\documentclass[%
reprint,
superscriptaddress,
showpacs,preprintnumbers,
bibnotes,
amsmath,amssymb,
aps,
]{revtex4-1}

\usepackage{graphicx}
\usepackage{dcolumn}
\usepackage{bm}
\usepackage{xcolor}

\begin{document}

\newcommand{\BTS}{Bi$_2$Te$_2$Se}
\newcommand{\Tm}{$T_m$}
\newcommand{\Ti}{$T_i$}
\newcommand{\Hon}{$H_{\rm onset}$}
\newcommand{\cvs}{cm$^2$V$^{-1}$s$^{-1}$}
\newcommand{\Ea}{$E_a$}
\newcommand{\ie}{{\it i.e.}}
\newcommand{\eg}{{\it e.g.}}
\newcommand{\etal}{{\it et al.}}  
\newcommand{\mucm}{$\mathrm{\mu\Omega\, cm}$}
\newcommand{\Hcstar}{$H_c2^\star$}
\newcommand{\RH}{$R_H$}


\title{Temperature-field phase diagram of extreme magnetoresistance in lanthanum monopnictides}

\author{F.~F.~Tafti}
\email{ftafti@princeton.edu}
\affiliation{Princeton University, Chemistry department, Princeton, New Jersey, USA}

\author{Q.~D.~Gibson}
\affiliation{Princeton University, Chemistry department, Princeton, New Jersey, USA}

\author{S.~K.~Kushwaha}%
\affiliation{Princeton University, Chemistry department, Princeton, New Jersey, USA}

\author{J.~W.~Krizan}
\affiliation{Princeton University, Chemistry department, Princeton, New Jersey, USA}

\author{N.~Haldolaarachchige}
\affiliation{Princeton University, Chemistry department, Princeton, New Jersey, USA}

\author{R.~J.~Cava}
\affiliation{Princeton University, Chemistry department, Princeton, New Jersey, USA}
\email{rcava@princeton.edu}

\date{\today}

\begin{abstract}
The recent discovery of extreme magnetoresistance in LaSb introduced lanthanum monopnictides as a new platform to study topological semimetals (TSMs).
In this work we report the discovery of extreme magnetoresistance in LaBi confirming lanthanum monopnictides as a promising family of TSMs. 
These binary compounds with the simple rock-salt structure are ideal model systems to search for the origin of extreme magnetoresistance.
Through a comparative study of magnetotransport effects in LaBi and LaSb, we construct a \emph{triangular} temperature-field phase diagram that illustrates how a magnetic field tunes the electronic behavior in these materials.
We show that the triangular phase diagram can be generalized to other topological semimetals with different crystal structures and different chemical compositions.
By comparing our experimental results to band structure calculations, we suggest that extreme magnetoresistance in LaBi and LaSb originates from a particular orbital texture on their qasi-2D Fermi surfaces.
The orbital texture, driven by spin-orbit coupling, is likely to be a generic feature of various topological semimetals.
\end{abstract}

\pacs{75.47.-m, 71.70.Ej, 71.30.+h, 71.18.+y}
\maketitle


\section{\label{introduction}Introduction}

%
Materials with large magnetoresistance (MR) have applications in electronics as magnetic memories  \cite{daughton_gmr_1999, rao_giant_1996}, in spintronics as magnetic valves \cite{wolf_spintronics:_2001}, and in industry as magnetic sensors or magnetic  switches \cite{lenz_review_1990, jankowski_hall_2011}. 
The magnitude of MR is determined by the ratio  $R(H)/R(0)$ where $R$ is the electrical resistance and $H$ is the magnetic field.
This ratio is large either when the numerator $R(H)$ is large or when the denominator $R(0)$ is small.
A large numerator is responsible for the well-known Giant and Collosal Magnetoreistance (GMR and CMR) in magnetic semiconductors \cite{rao_giant_1996, ramirez_colossal_1997-1}.
A small denominator is reponsible for the recently-found extreme magnetoresistance (XMR) in Dirac semimetals (DSMs) such as Na$_3$Bi or Cd$_3$As$_2$ \cite{xiong_evidence_2015, liang_ultrahigh_2015}, in Weyl semimetals (WSMs) such as NbP, NbAs, or TaAs \cite{shekhar_extremely_2015, ghimire_magnetotransport_2015, huang_observation_2015}, and in layered semimetals (LSMs) such as WTe$_2$, NbSb$_2$, or PtSn$_4$ \cite{ali_large_2014, ali_correlation_2015, zhu_quantum_2015, wang_anisotropic_2014, mun_magnetic_2012}.
DSMs are characterized by strictly linear band crossing at the Dirac point. \cite{borisenko_experimental_2014}.
WSMs are characterized by noncentrosymmetric structures, Fermi arcs, and chiral anomaly \cite{xiong_evidence_2015, weng_weyl_2015, xu_discovery_2015_NbAs}.
LSMs are characterized by nearly perfect compensation between electron and hole bands \cite{ali_large_2014, wang_anisotropic_2014, pletikosic_electronic_2014}.
The recent discovery of XMR in LaSb that lacks strictly linear band crossing, non-centrosymmetric structure, chiral anomaly, and perfect compensation shows that none of these factors control XMR.
This work presents a systematic study to understand the origin of XMR. 

\begin{figure}
\includegraphics[width=3.5in]{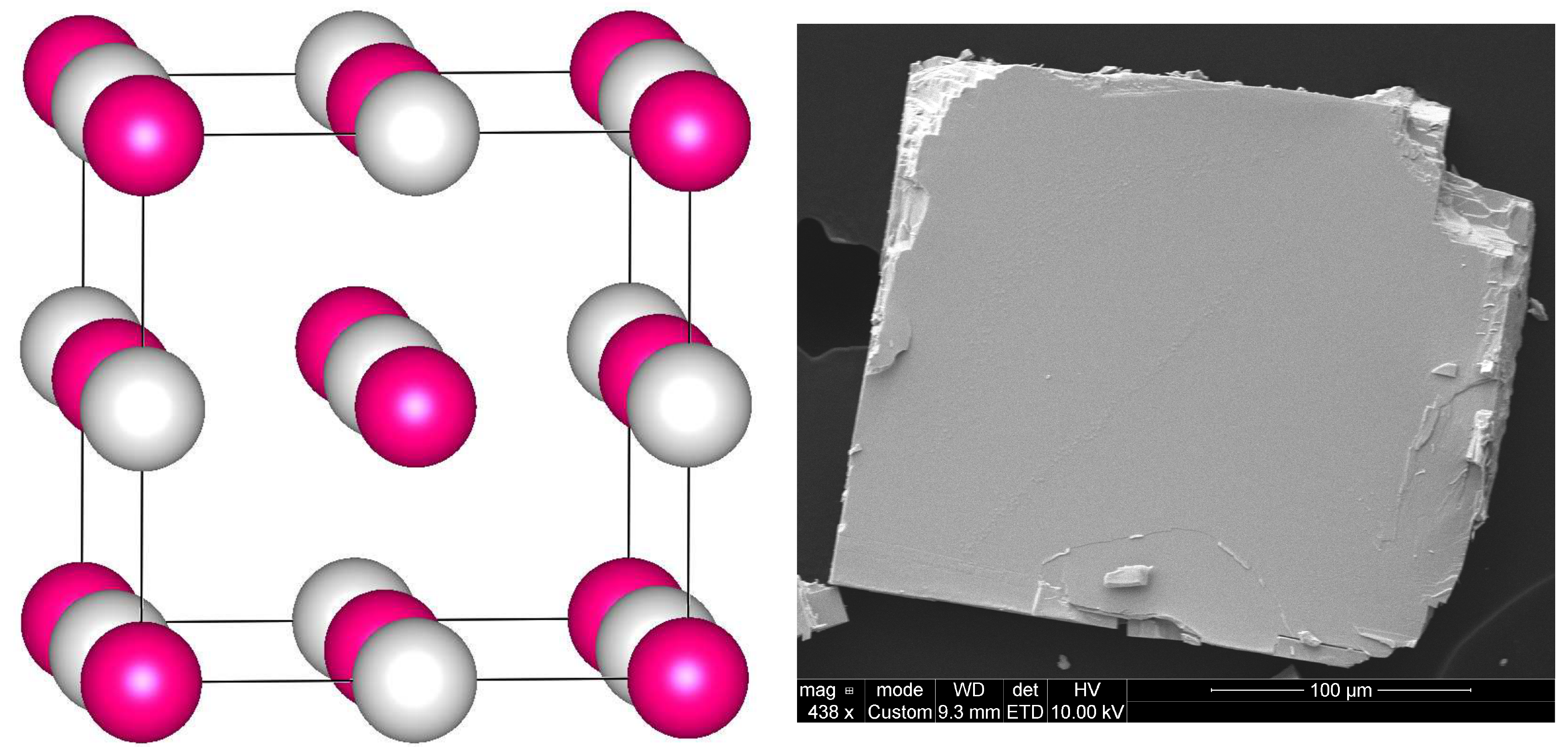}
\caption{\label{Structure} 
(left) NaCl-type crystal structure of LaSb/LaBi with space group $Fm\bar{3}m$. 
Gray spheres represent La and pink spheres represent Sb/Bi.
(right) Environmental Scanning Electron Microscopy (ESEM) image of a single crystal of LaBi showing its cubic symmetry and the cleaved [100] surface of the crystal. 
X-ray patterns and structural refinements are presented in Appendix A.
}
\end{figure}

We report the discovery of XMR in LaBi with similar crystal structure and chemical composition to LaSb.
A recent theoretical work has pointed out that the simple NaCl-type structure of lanthanum monopnictides (Fig. \ref{Structure}) makes them ideal model systems to study the properties of topological semimetals \cite{zeng_topological_2015}.
Through a comparative study of longitudinal and transverse magnetotransport effects in the two related materials, LaSb and LaBi, we construct a \emph{triangular} $T$-$H$ phase diagram for XMR in lanthanum monopnictides.
From band structure calculations, quantum oscillations, and Hall Effect measurements, we suggest that XMR is rooted in a $d$-$p$ orbital mixing on the quasi-2D electron pockets that dominate the low temperature electrical transport in lanthanum monopnictides. 
Further, we show that similar quasi-2D Fermi surfaces with mixed orbital content exist in other TSMs and so does the triangular $T$-$H$ phase diagram.
These results provide compelling evidence for XMR being rooted in quasi-2D Fermi surfaces with mixed $d$-$p$ orbital texture.
%
%
%
%
%
%
%
%
%


\section{\label{methods}Methods}

Single crystals of LaBi were grown using Indium flux.
The starting elements La:Bi:In = 1:1:20 with purity $99.999\%$ were placed in an alumina crucible inside an evacuated quartz tube.
The mixture was heated to 1000 C, slowly cooled to 700 C, and finally decanted in a centrifuge. 
Similar procedure was used to grow single crystals of LaSb from tin flux \cite{tafti_consequences_2015}.
Energy dispersive x-ray spectroscopy (EDX) on each sample confirmed a 1:1 ratio of lanthanum to pnictogen with $\pm 1 \%$ error.
X-ray diffraction (XRD) patterns of both materials are presented in Appendix A.
Measurements were performed on two samples: one single crystal of LaSb with residual resistivity $\rho_0 = 0.6$ \mucm~and residual resistivity ratio RRR $= \rho(300 {\textrm K}) / \rho_0=$ 170 and one single crystal of LaBi with $\rho_0 = 0.1$ \mucm~and RRR = 610.
%
%
The resistivity measurements were performed in a Quantum Design PPMS using a standard four probe method.
Hall voltages were measured with transverse contacts in negative and positive fields with the data antisymmetrized to calculate the transverse resistivity $\rho_{xy}$ and the Hall coefficient \RH $=\rho_{xy} / H$. 

\section{\label{results}Results}

\subsection{\label{temperature}Temperature dependence of resistivity}

\begin{figure}
\includegraphics[width=3.5in]{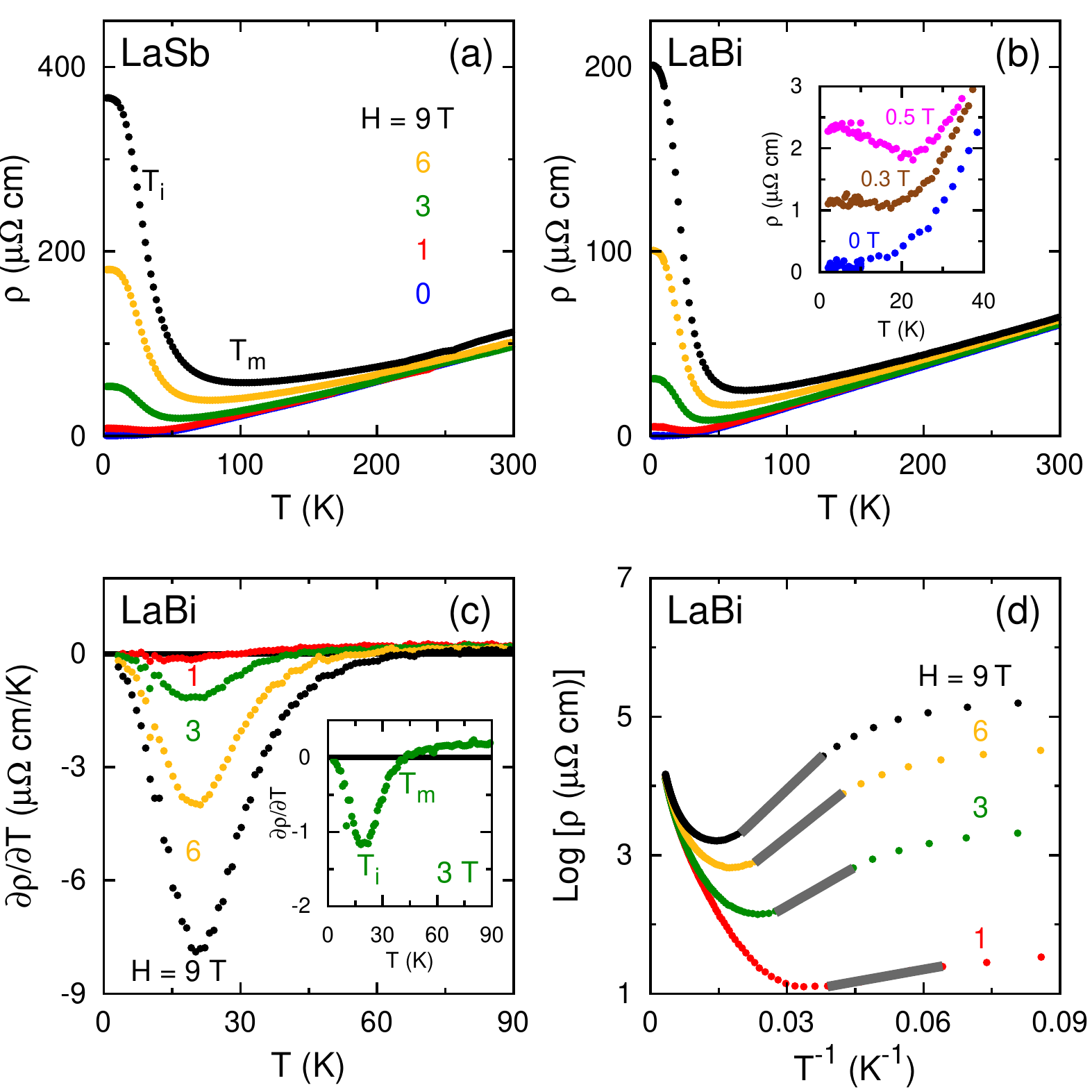}
\caption{\label{RT} 
(a) Resistivity as a function of temperature in LaSb at several magnetic fields as indicated on the figure. 
Temperature of resistivity minimum \Tm~and inflection \Ti~are marked on the black curve at $H=9$ T.
(b) $\rho(T)$ in LaBi at the same fields as in LaSb.
Inset shows $\rho(T)$ at low $T$ and low $H$ to capture the onset of resistivity activation at $H_{\rm onset} = 0.4\pm 0.1$ T.
(c) $\partial \rho / \partial T$ as a function of temperature in several fields as indicated on the figure for LaBi.
Inset shows the sign change temperature \Tm~and the peak temperature \Ti~for the green curve at $H = 3$ T.
(d) Arrhenius plot of $\log(\rho)$ versus $T^{-1}$ used to extract the activation energy \Ea~at several fields as indicated on the figure for LaBi.
}
\end{figure}

Fig. \ref{RT}(a) and \ref{RT}(b) show the temperature dependence of resistivity in LaSb and LaBi.
In both systems, with decreasing temperature, resistivity decreases initially until a minimum at \Tm, then increases until an inflection at \Ti~where it gradually saturates to a plateau.
Fig. \ref{RT}(c) is a plot of $\partial \rho / \partial T$ versus $T$ for LaBi which marks \Tm~as the sign change temperature and \Ti~as the peak temperature.
With increasing field, \Tm~increases but \Ti~remains unchanged.
Fig. \ref{RT}(d) is an Arrhenius plot of $\log(\rho)$ versus $T^{-1}$ for LaBi.
It shows that between \Tm~and \Ti~the material behaves like semiconductors with $\rho(T) \propto \exp(E_a/k_{\rm B}T)$ where \Ea~is an activation energy and $k_{\rm B}$ is the Boltzmann constant.
The activation energy at each field corresponds to the slope of the linear fits in Fig. \ref{RT}(d)
The LaSb versions of Figs. \ref{RT}(c) and \ref{RT}(d) are presented in Ref. \cite{tafti_consequences_2015}.
Inset of Fig.~\ref{RT}(b) shows that the resistivity activation in LaBi is absent at zero field; it is switched on with a small magnetic field \Hon~$= 0.4\pm 0.1$ T similar to LaSb \cite{tafti_consequences_2015}.
Both materials have potential application as low temperature magnetic switches.
Below \Tm, $\rho(T)$ in LaSb and LaBi is analogous to $\rho(T)$ in topological insulators (TIs) such as Bi$_2$Te$_2$Se and SmB$_6$.
In TIs the resistivity activation comes from an insulating bulk, and the plateau from conducting surface states \cite{ren_large_2010, jia_defects_2012, kim_surface_2013, kim_topological_2014}.
In LaSb and LaBi, a similar activation and plateau appear only when time reversal symmetry is broken by a small magnetic field.
In section \ref{angular}, we show that the conduction in the plateau region of these materials is dominated by quasi-2D bulk states in analogy to the strictly 2D surface states of TIs.

\subsection{\label{phase}Phase Diagram}

\begin{figure}
\includegraphics[width=3.5in]{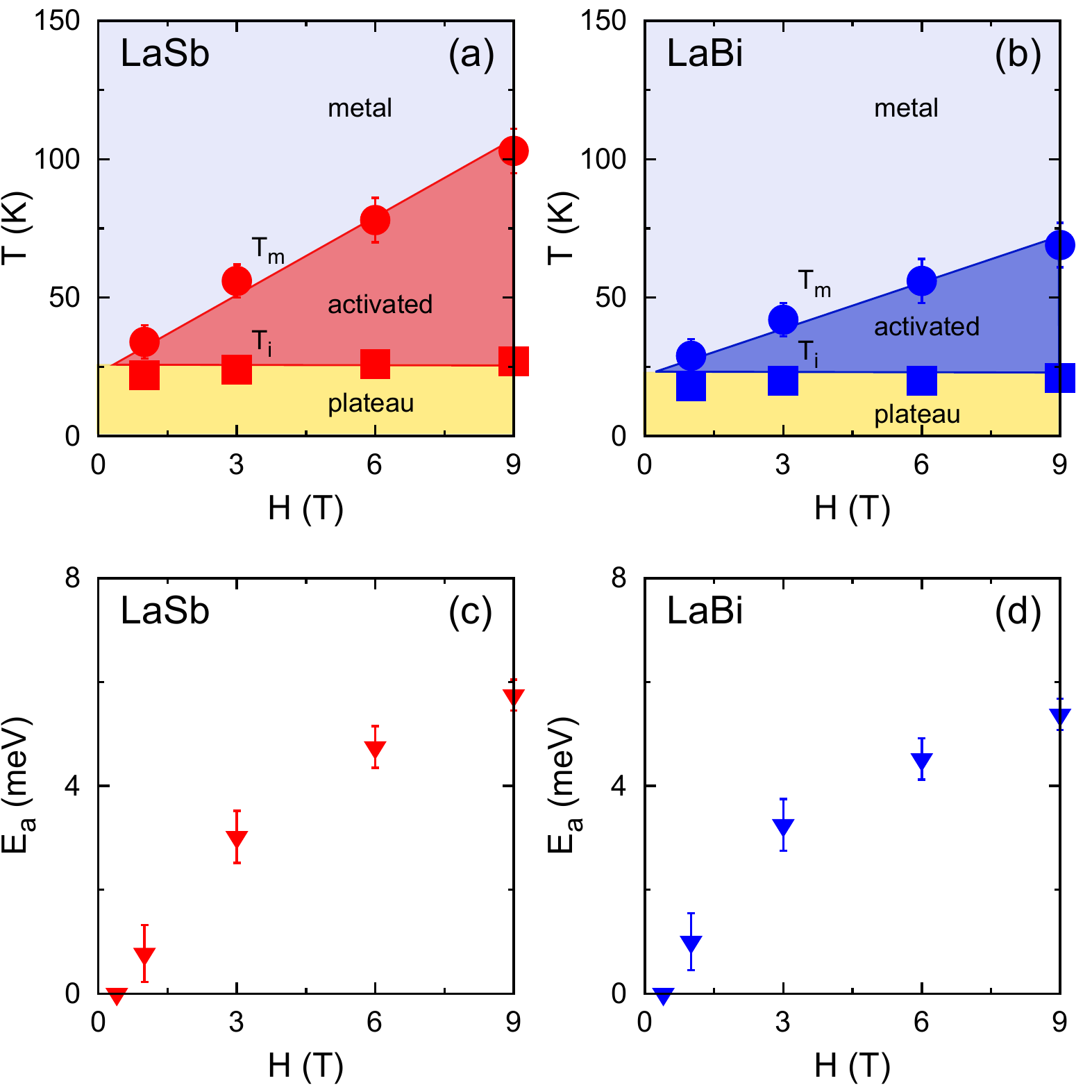}
\caption{\label{PD} 
(a) \Tm~(circles) and \Ti~(squares) plotted as a function of $H$ in LaSb.
The two temperature scales merge at \Hon~$\simeq 0.4$ T and diverge with increasing field giving rise to a triangle.
In the triangular region, $\rho(T)$ shows field-induced activation.
The silver region above the triangle is where $\rho(T)$ behaves like a metal.
The gold region below the triangle is where $\rho(T)$ plateaus.
(b) Similar triangular phase diagram for LaBi.
(c) Activation energy \Ea~extracted from the Arrhenius analysis in the shaded triangular region and plotted as a function of magnetic field for LaSb.
(d) \Ea~plotted as a function of field for LaBi with a behavior comparable to LaSb.
}
\end{figure}

By plotting \Tm~and \Ti~as a function of magnetic field, we construct the temperature-field phase diagram of LaSb and LaBi in Fig.~\ref{PD}(a) and \ref{PD}(b). 
In both systems, \Tm~increases with increasing field while \Ti~stays unchanged.
The shaded triangle between \Tm~and \Ti~marks the region of activated resistivity.
Fig. \ref{PD}(c) and \ref{PD}(d) show that the activation energy \Ea~starts from zero at \Hon~and increases with a non-monotonic field dependence.
\Ea~at each field is extracted from the Arrhenius analysis explained in section \ref{temperature} and Fig. \ref{RT}(d).
Notably \Ea~is comparable in both systems and scales only with the magnetic field showing that the resistivity activation is controlled entirely by the magnetic field. 
The triangular phase diagrams in Figs. \ref{PD}(a) and \ref{PD}(b) highlight three distinct resistivity behaviors in Figs. \ref{RT}(a) and \ref{RT}(b).
The shaded triangle is the region of activated resistivity with semiconducting behavior.
In the silver region above the triangle ($T>$ \Tm) the semiconducting behavior is replaced with the metallic conduction. 
In the gold region below the triangle ($T<$ \Tm) the semiconduting behavior is replaced with the plateau. 
\Tm~and \Ti~merge at \Hon~and diverge as the field is increased. 
In section \ref{universal} we show that a similar phase diagram can be constructed from the existing data in several transition-metal-based TSMs.

\subsection{\label{extreme}Field dependence of resistivity and extreme magnetoresistance}

\begin{figure}
\includegraphics[width=3.5in]{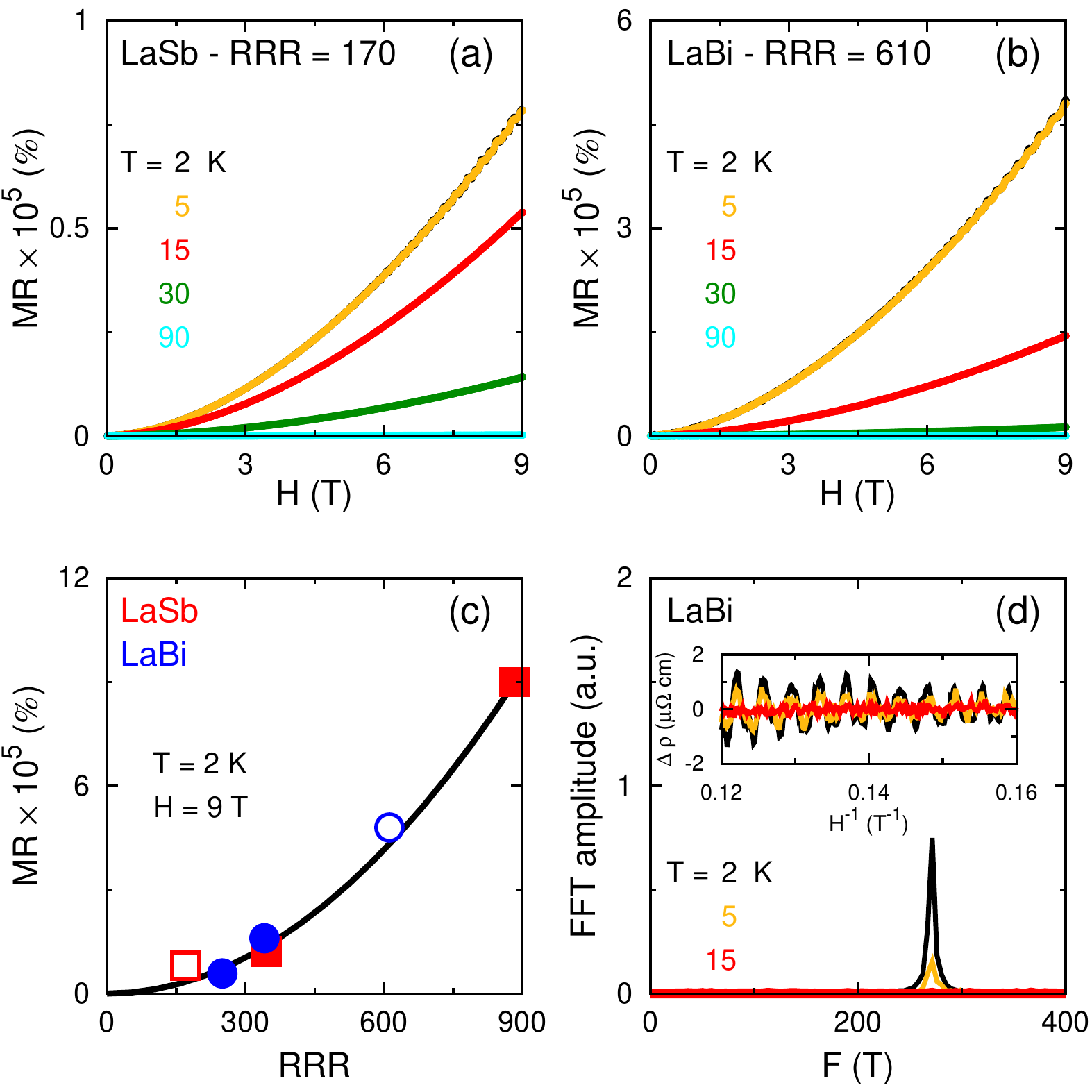}
\caption{\label{RH} 
(a) Magnetoresistance as a function of field in LaSb sample with RRR = 170  at several temperatures indicated on the figure.
SdH oscillations appear at higher fields.
(b) MR as a function of field for LaBi sample with RRR = 610.
(c) MR at $T = 2$ K and $H = 9$ T plotted as a function of RRR for several LaSb (square) and LaBi (circle) samples from this work and Ref. \cite{tafti_consequences_2015}. 
Empty symbols mark the two samples studied here in detail.
(d) Inset shows the oscillatory part of resistivity, $\Delta \rho (H^{-1})$, at several temperatures for LaBi.
Fast Fourier Transform of these data are plotted as a function of frequency with a peak at $F_0=217$~T.
}
\end{figure}

%
%
Figs. \ref{RH}(a) and \ref{RH}(b) show magnetoresistance, MR = $100\times[R(H)-R(0)]/R(0)$, as a function of magnetic field from $H=0$ to 9 T at several temperatures. 
The black and the yellow curves at $T$ = 2 K and 5 K superimpose since both curves are in the plateau region where MR reaches its \emph{extreme} limit in excess of $10^5 \%$.
Comparing Figs. \ref{RH}(a,b) with the phase diagram Figs. \ref{PD}(a,b) shows that MR is small in the bulk metallic phase at $T>T_m$, it starts to increase in the semiconducting phase at $T_m>T>T_i$, and reaches the extreme limit in the plateau region at $T<T_i$.  
Ref. \cite{tafti_consequences_2015} shows that the magnitude of XMR is sensitive to the residual resistivity ratio RRR \ie~to the sample quality.
We quote RRR values for the LaSb and the LaBi samples in Fig. \ref{RH}(a) and \ref{RH}(b) to prevent the illusion that XMR in LaSb is smaller than LaBi.
Fig. \ref{RH}(c) is a plot of MR as a function of RRR for several LaSb and LaBi specimens with different RRR values.
Empty symbols mark the two samples presented in this work. 
MR in both materials follows the same quadratic dependence on RRR showing a comparable XMR in both compounds given comparable sample quality. 
A similar quadratic dependence of XMR on RRR is reported in the flux-grown WTe$_2$ samples \citep{ali_correlation_2015}.

%
%

The ripples at higher fields in Figs. \ref{RH}(a) and \ref{RH}(b) are Shubnikov-de Haas (SdH) oscillations.
The purely oscillatory part of resistivity $\Delta\rho (H)$ is obtained by subtracting a smooth $H^2$ background from $\rho(H)$.
$\Delta \rho$ is periodic in $H^{-1}$ as seen in the inset of Fig. \ref{RH}(d).
Fast Fourier Transform (FFT) of these data gives a peak at $F_0=271 \pm 5$ T in Fig. \ref{RH}(d) for LaBi.
Similar analysis gives $F_0=212 \pm 5$ T for LaSb \cite{tafti_consequences_2015}.
The principal frequencies for LaSb and LaBi do not match the known frequencies of tin and indium ruling out flux inclusion \cite{deacon_ultrasonic_1973, cowey_quantum_1974}.

\subsection{\label{angular}Angle dependence of XMR and SdH oscillations}

\begin{figure}
\includegraphics[width=3.5in]{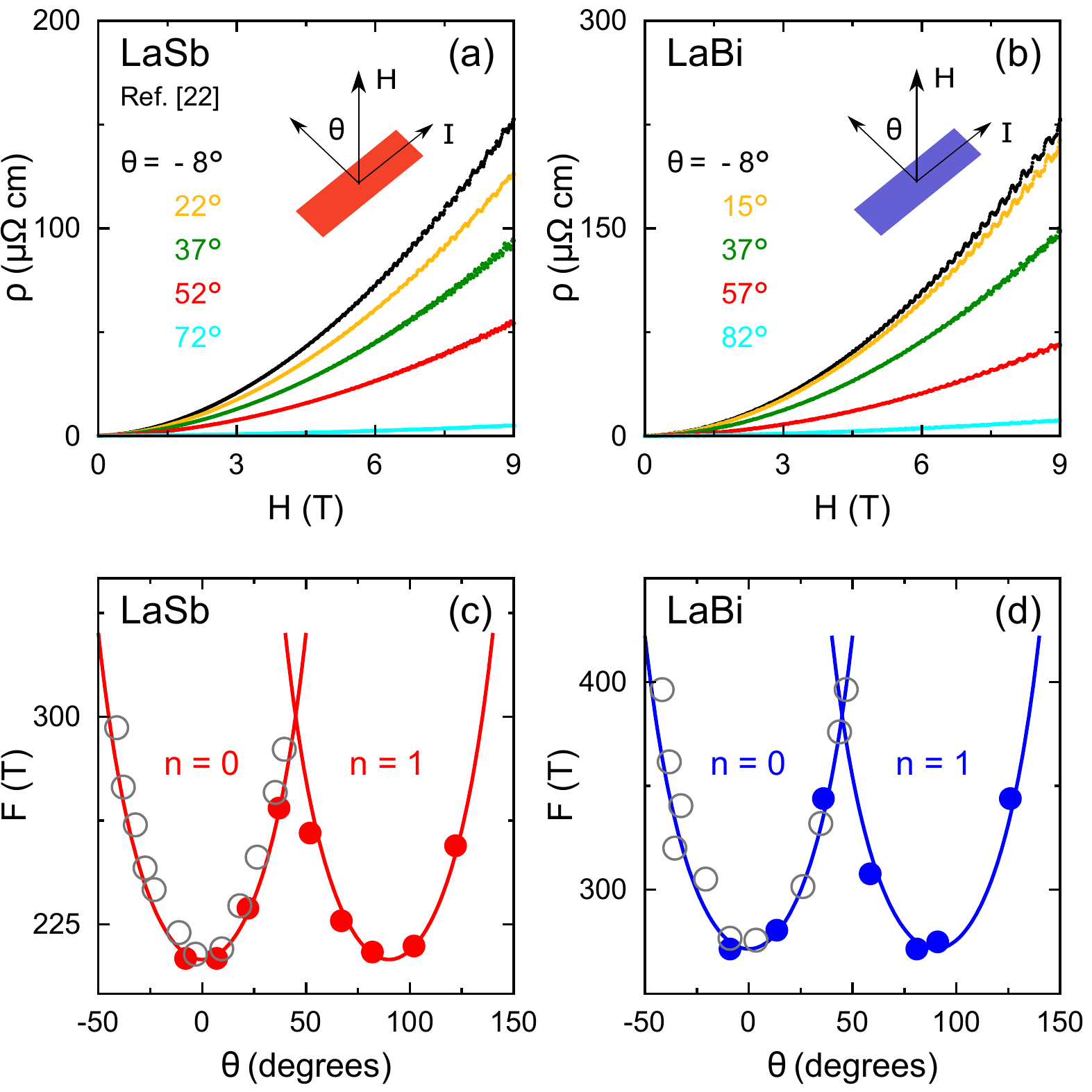}
\caption{\label{Angle} 
(a) Resistivity as a function of magnetic field in LaSb at five representative angles as indicated.   
MR decreases with increasing $\theta$.
%
Shubnikov - de Haas (SdH) frequencies vary non-monotonically with $\theta$. 
LaSb data are adapted from Ref. \cite{tafti_consequences_2015}.
(b) $\rho(H)$ in LaBi at five representative angles as indicated.
MR stays positive at all angles in both systems.
(c) SdH frequency as a function of angle in LaSb.
Empty gray circles are de Haas-van Alphen (dHvA) data from Ref. \cite{kitazawa_haas-van_1983} limited to lower angles.
Solid lines are fits to Eq. \ref{cos} with $F_0 = 212$ T. 
(d) SdH oscillation frequency as a function of angle in LaBi.
Solid lines are fits to Eq. \ref{cos} with $F_0 = 271$ T. 
Empty gray symbols are dHvA data from Ref. \cite{hasegawa_fermi_1985}.
}
\end{figure}

Figs. \ref{Angle}(a) and \ref{Angle}(b) show the angle dependence of XMR in LaSb and LaBi.
The direction of magnetic field $H$ and electrical current $I$ with respect to [100] crystal plane are shown schematically.
XMR is maximum when $H\perp I$ and minimum when $H\| I$.
It remains positive at all angles.
Figs. \ref{Angle}(c) and \ref{Angle}(d) show that the principal frequency in LaSb and LaBi follows the angle dependence of a two dimensional Fermi surface:
\begin{equation} 
F = \frac{F_0}{\cos \left( \theta - \frac{n \pi}{2} \right)} 
\label{cos}
\end{equation}
where $n$ is an integer, $\theta$ is the angle, and $F_0$ is the principal frequency at $\theta = 0$.
These data show that the main frequencies at $F_0 = 212$ T in LaSb and $F_0 = 271$ T in LaBi belong to quasi-2D Fermi surfaces.
Fig. \ref{Angle}(c) shows a good agreement between our data and prior studies of magnetic oscillations (gray symbols) \cite{kitazawa_haas-van_1983, settai_acoustic_1993, yoshida_cyclotron_2001, hasegawa_fermi_1985}.
Similar angle dependence of quantum oscillations in topological insulators has been taken as evidence for strictly 2D surface states of TIs \cite{ren_large_2010, li_two-dimensional_2014, tan_unconventional_2015}.
SdH frequencies of LaSb and LaBi shown in Fig. \ref{Angle} come from quasi-2D bulk states and not from strictly 2D surface states (section \ref{bandstructure}). 
As pointed out in Ref. \cite{tafti_consequences_2015}, the combination of 2D or quasi-2D Fermiology and band inversion gives rise to analogous transport phenomenology in TIs and TSMs.
%

\subsection{\label{bandstructure}Band structure and the orbital texture of the quasi-2D Fermi surface}

\begin{figure}
\includegraphics[width=3.5in]{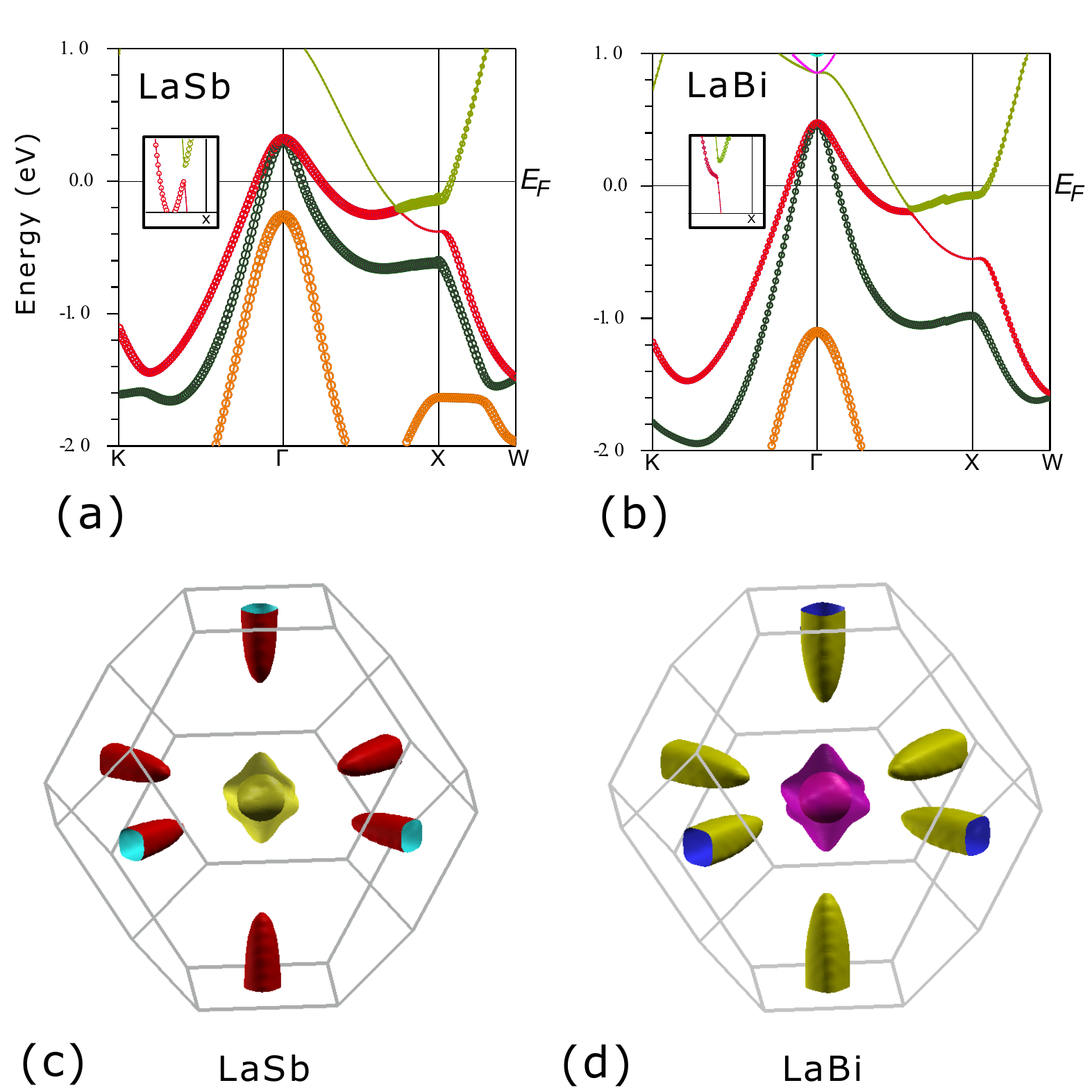}
\caption{\label{Calc} 
(a) Band structure of LaSb with two hole pockets that cross $E_F$ near $\Gamma$ and one electron pocket that crosses $E_F$ near $X$.
The crossing between lanthanum $d$-states (thin bands) and Antimony $p$-states (thick bands) near $X$ gives a $d$-$p$ texture to the electron pocket due to spin-orbit coupling. 
The inset shows a small gap at this crossing.
(b) Band structure of LaBi with the same $d$-$p$ crossing between La thin bands and Bi thick bands near $X$.
(c) Fermi surfaces of LaSb visualized in the first Brillouin zone with the central 3D hole pockets and the peripheral quasi-2D electron pockets.
SdH frequencies in Fig. \ref{Angle} correspond to these quasi-2D electron pockets.
(d) Similar Fermi surfaces for LaBi.
}
\end{figure}

Figs. \ref{Calc}(a) and \ref{Calc}(b) show the results of our band structure calculations on LaSb and LaBi using the WIEN2k code \cite{blaha_wien2k_2001}.
In both systems, two hole bands at the $\Gamma$-point and one electron band near the $X$-point cross $E_F$.
Figs. \ref{Calc}(c) and \ref{Calc}(d) visualize the corresponding Fermi surfaces with the two hole surfaces at the center of the Brillouin zone and the quasi-2D electron pockets crossing the faces. 
Prior studies of quantum oscillations in the magnetic and acoustic channels have detected these hole and electron Fermi surfaces in both materials \cite{kitazawa_haas-van_1983, settai_acoustic_1993, yoshida_cyclotron_2001, hasegawa_fermi_1985}.
The principal SdH frequencies that we observe in LaSb ($F_0=212$ T) and LaBi ($F_0=271$ T) match the cross-sectional area of the quasi-2D surface at the X-point.
Therefore, electrical transport in the plateau region of these materials is dominated by the quasi-2D electron pocket.
We plot Antimony $p$-states as thick bands and lanthanum $d$-states as thin bands in Figs. \ref{Calc}(a) and \ref{Calc}(b).
These states cross near the $X$-point and form the quasi-2D electron pocket with a mixed $d$-$p$ orbital texture.
The inset in both Figures show a small gap at the crossing point driven by strong spin-orbit coupling similar to topological insulators.
%
%
%
Therefore, the quasi-2D pocket that dominates the low temperature transport could acquire topological protection against scattering similar to the surface states of TIs.
A magnetic field could interfere with the $d$-$p$ orbital mixing and activate strong scattering on these pockets giving rise to XMR.
Recent observations of circular dichroism by ARPES confirms the same orbital mixing in WTe$_2$ \cite{jiang_signature_2015}.
In Appendix \ref{DFTs} we show that the band structure of WTe$_2$ and several other topological semimetals have the same orbital texture as lanthanum monopnictides (Fig. \ref{Calc_All}). 
In section \ref{universal}, we show that these other TSMs also have the same triangular phase diagram as lanthanum monopictides.
These observations strongly suggest that XMR in various TSMs is a result of orbital texture on quasi-2D Fermi surfaces.

\subsection{\label{SdH}Effective mass and Dingle temperature}

\begin{figure}
\includegraphics[width=3.5in]{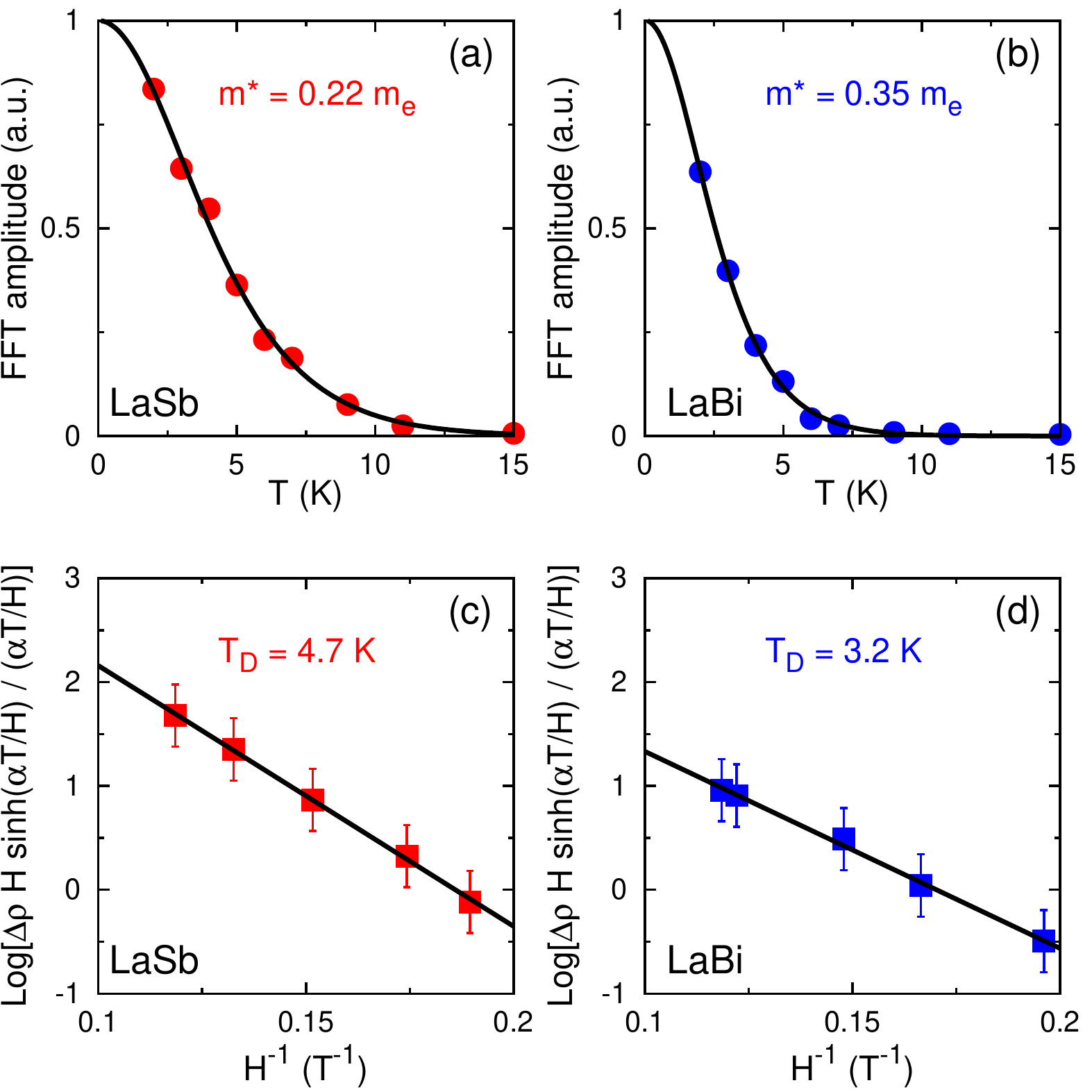}
\caption{\label{QO} 
(a) Amplitude of the Fast Fourier transform (FFT) of SdH oscillations with $F_0=212$ T plotted as a function of temperature for LaSb.
Solid line is a fit to the Lifshitz-Kosevich formula (Eq. \ref{lifshitz}).  
The effective mass of carriers is extracted from the fit and reported on the figure.
(b) Lifshitz-Kosevich analysis for LaBi.
Effective mass in LaBi is slightly larger than LaSb.
(c) FFT amplitudes modulated by the Lifshitz-Kosevich factor plotted as a function of $H^{-1}$ for LaSb.
Solid line is a fit to the Dingle formula (Eq. \ref{dingle}).
Dingle temperature $T_D$ is reported on the figure. 
(d) Dingle analysis for LaBi.
}
\end{figure}

Using the Onsager relation, we extract the Fermi wave vector and the density of carriers on the quasi-2D electron pocket from the frequency of SdH oscillations: 
\begin{equation}
F = \frac{\phi_0}{2\pi ^2} A_{\rm ext} ~,~ A_{\rm ext} = \pi k_F^2 ~,~ k_F = \left(4 \pi n_{2D} \right)^{1/2} 
\label{onsager}
\end{equation}
where $\phi_0$, $A_{\rm ext}$, $k_F$, and $n_{2D}$ are the quantum of flux, the extremal orbit area, the Fermi wave vector, and the two dimensional carrier density for spin filtered electrons.  
Fig. \ref{RH}(d) shows $F = 271$ T in LaBi, therefore, $k_{\rm F} = 9.1\times 10^6$ cm$^{-1}$ and $n_{2D} = 6.6 \times 10^{12}$ cm$^{-2}$.
Corresponding values for LaSb are given in Table \ref{T1}.
The oscillation amplitude damps with increasing temperature and with decreasing magnetic field (Fig. \ref{RH}(d)) according to:
\begin{equation}
\Delta \rho = R_L R_D \sin \left( \frac{2 \pi F}{H} + \phi \right) 
\label{damp}
\end{equation}
$R_L$ is the Lifshitz-Kosevich factor that captures damping with increasing temperature: 
\begin{equation}
R_{L} = \frac{X}{\sinh(X)} \qquad , \qquad X = \frac{\alpha T m^*}{H} 
\label{lifshitz}
\end{equation}
where $\alpha = 2 \pi ^2 k_B m_e / e \hbar$ is a constant made of Boltzmann factor $k_B$, bare electron mass $m_e$, electron charge $e$, and reduced Plank constant $\hbar$.
$m^*$ is the effective electron mass in units of $m_e$.
$R_D$ is the Dingle factor that captures damping with decreasing field:
\begin{equation}
R_D = \exp(-X_D) \qquad , \qquad X_D = \frac{\alpha T_D m^*}{H} 
\label{dingle}
\end{equation}
where $T_D$ is the Dingle temperature from which the relaxation rate $\tau$, the mean free path $\ell$, and the mobility $\mu$ of charge carriers can be determined using:
\begin{equation}
k_B T_{D} = \frac{\hbar}{2 \pi \tau} \quad , \quad \ell = v_F \tau \quad , \quad \mu = \frac{e \tau}{m^*}
\label{mfp}
\end{equation}
with the Fermi velocity $v_F = \hbar k_F / m^*$. 
Figs. \ref{QO}(a) and \ref{QO}(b) show the Lifshitz-Kosevich fit (Eq. \ref{lifshitz}) to the temperature dependence of the oscillation amplitude.
The resulting effective masses are $m^*=0.22\,m_e$ for LaSb and $m^*=0.35\,m_e$ for LaBi.
Figs. \ref{QO}(c) and \ref{QO}(d) show the Dingle fit (Eq. \ref{dingle}) to the field dependence of the oscillation amplitude that determines $\tau$, $\ell$, and $\mu$ for the carriers in the plateau region of LaSb and LaBi.
Table \ref{T1} summarizes all the parameters from SdH oscillations and compares them to the corresponding values in the topological insulator \BTS~\cite{ren_large_2010}.

\begin{table*}
\caption{\label{T1} 
Fermi surface parameters are tabulated for LaSb, LaBi, and \BTS~from the analysis explained in section \ref{SdH} (Eqs. \ref{onsager} to \ref{mfp}). Data for \BTS~are adapted from Ref. \cite{ren_large_2010}.
}
\begin{ruledtabular}
\begin{tabular}{cccccccccc}
Material & $F$ & $m^*/m_e$ & $T_{\rm D}$ & $k_{\rm F}$ & $n_{\rm 2D}$ & $v_{\rm F}$ & $\tau$ & $\ell$ & $\mu_e$ 
\\
         & [T] &                  &     [K]     & [cm$^{-1}$] &  [cm$^{-2}$] & [cm s$^{-1}$]&  [s]   &  [nm]  & [cm$^2$V$^{-1}$s$^{-1}$]
\\       
\colrule 
LaSb & $212$ & $0.22$ & $7.8$ & $8.0 \times 10^6$ & $5.1 \times 10^{12}$ & $4.2 \times 10^7$ & $1.6 \times 10^{-13}$ & 66 & 1250 
\\
LaBi & $271$ & $0.35$ & $3.7$ & $9.1 \times 10^6$ & $6.6 \times 10^{12}$ & $3.0 \times 10^7$ & $3.3 \times 10^{-13}$ & 98 & 1650 
\\
\BTS & 64 & 0.1 & 25.5 & $4.4 \times 10^6$ & $1.5 \times 10^{12}$ & $4.6 \times 10^7$ & $4.8 \times 10^{-14}$ & 22 & 760 

\end{tabular}
\end{ruledtabular}
\end{table*}

\subsection{\label{halleffect}Hall effect and the two band model}

\begin{figure}
\includegraphics[width=3.5in]{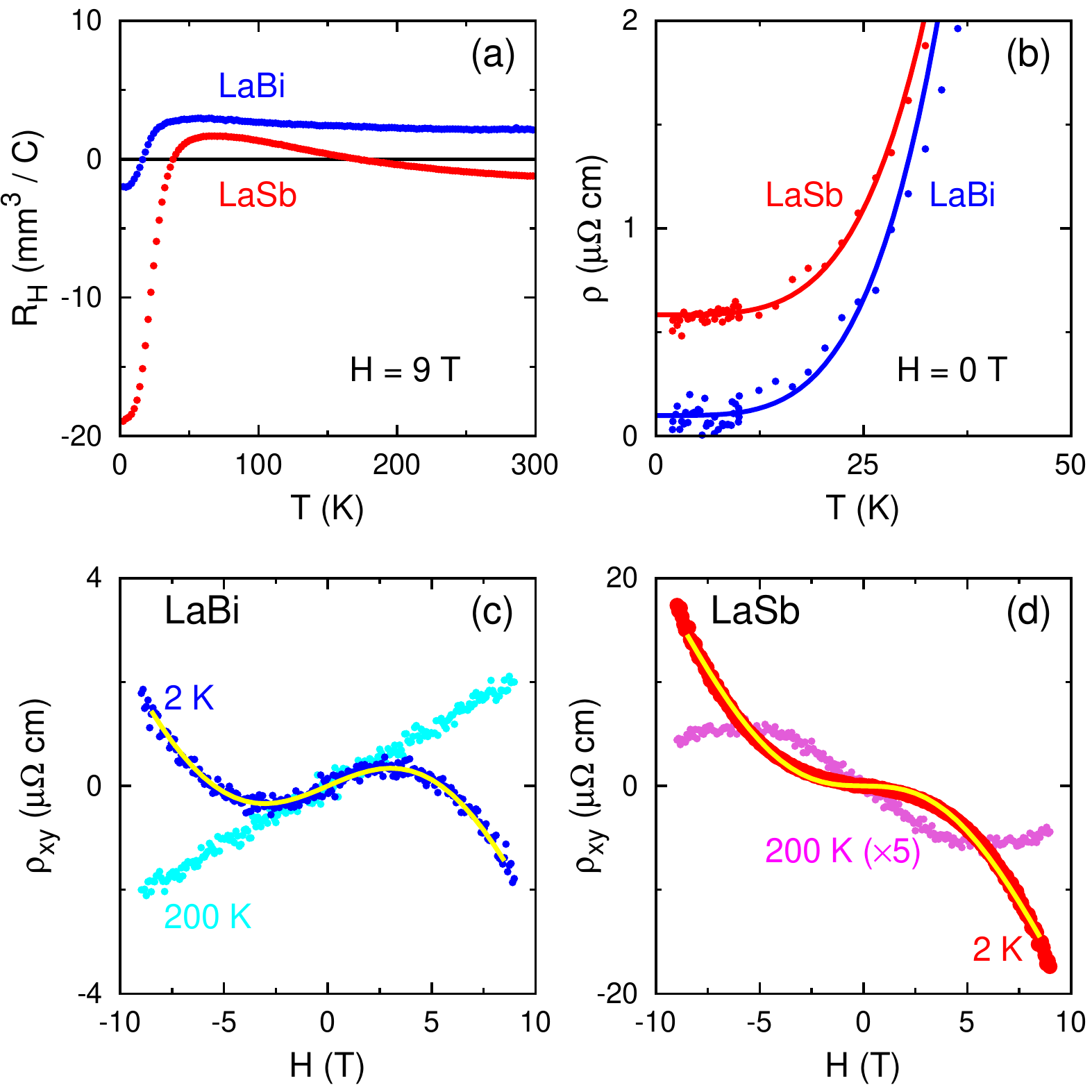}
\caption{\label{RhT} 
(a) $R_H$ at $H=9$ T plotted as a function of temperature from $T = 2$ to 300 K in LaSb (red) and LaBi (blue).
There is one sign change in LaBi at $T \simeq 20$ K and two sign changes in LaSb at $T \simeq 40$ K and 170 K.   
(b) $\rho(T)$ at $H=0$ T and $T<40$ K in LaSb and LaBi.
Solid lines are fits of the form $\rho = \rho_0 + AT^4$ to determine $\rho_0=0.6$ \mucm~in LaSb and 0.1 \mucm~in LaBi.
(c) $\rho_{xy}$ plotted as a function of field in LaBi from $H=-9$ to 9 T.
The blue curve at $T = 2$ K fits to the two band model Eq. \ref{twob} (solid yellow line).
The cyan curve at $T = 200$ K has a linear $H$ dependence.
(d) $\rho_{xy}$ plotted as a function of field in LaSb from $H=-9$ to 9 T.
The red curve at $T = 2$ K fits to the two band model Eq. \ref{twob} (solid yellow line).
The magenta curve at $T = 200$ K remains non-linear in $H$.
It is five times magnified for better visibility.
}
\end{figure}

Fig.~\ref{RhT}(a) shows temperature dependence of the Hall coefficient $R_H = \rho_{xy} / H$ at $H = 9$ T in LaSb (red) and LaBi (blue). 
LaBi shows a clear two band behavior with positive \RH~above $T \simeq 20$ K and negative \RH~of comparable magnitude below $T \simeq 20$ K. 
LaSb shows a strong negative \RH~signal below $T \simeq 40$ K and a weak positive signal above $T \simeq 40$ K that undergoes a second sign change at $T \simeq 170$ K.   
Fig. \ref{RhT}(b) shows power law fits to the resistivity data at low temperatures from which we extract residual resistivity $\rho_0=0.1$ \mucm~in LaBi and $\rho_0=0.6$ \mucm~in LaSb.
Using the $T=0$ limit of \RH~from Fig. \ref{RhT}(a) and $\rho_0$ from Fig. \ref{RhT}(b) we can estimate the transport mobility from the single band expression $\mu=R_H/\rho_0=3.2\times 10^4$ \cvs~in LaSb and $\mu=2.5\times 10^4$ \cvs~in LaBi. 
Single band estimates for TSMs need to be taken with caution due to their multi-band nature.
Note that these values are an order of magnitude larger than the more accurate values from quantum oscillations (table \ref{T1}).
Figs. \ref{RhT}(c) and \ref{RhT}(d) show the multiband behavior in the field dependence of the Hall resistivity $\rho_{xy}$ in LaBi and LaSb.
Solid yellow lines in both figures are fits to the two band expression for $\rho_{xy}$ at $T=2$ K \cite{takahashi_low-temperature_2011, xia_indications_2013}:
\begin{equation}
\rho_{xy} = \frac{H}{e}\frac{ ( n_h \mu_h^2 - n_e \mu_e^2 ) + ( n_h - n_e ) ( \mu_h \mu_e)^2 H^2}{ ( n_h  \mu_h + n_e \mu_e ) ^2 + ( n_h - n_e )^2 ( \mu_h \mu_e )^2 H^2} 
\label{twob}
\end{equation}
where $n_{e/h}$ and $\mu_{e/h}$ are the electron/hole carrier density and mobility.
Since the electron carriers at low temperatures come from the quasi-2D Fermi surfaces, we use the values in table \ref{T1} for $\mu_e$ and use $n_e = n^{2D}/d$ to calculate the effective 3D electron density where $d$ is the inter-layer spacing.
The fit gives an estimate for the hole carrier concentration $n_h$ and mobility $\mu_h$ which we summarize in table \ref{T2}.
The concentration of the hole carriers in LaBi agrees with a rough estimate using the single band formula $n_H = 1/eR_H = 3.1 \times 10^{21}$ cm$^{-3}$ using $R_H = 2$ mm$^3$C$^{-1}$ from Fig. \ref{RhT}(a).
This single band estimate in LaBi at high temperatures is justified by the linear field dependence of $\rho_{xy}$ at $T = 200$ K as seen in Fig. \ref{RhT}(c).
On the contrary, $\rho_{xy}$ in LaSb remains nonlinear at high temperatures as seen in Fig. \ref{RhT}(d) which explains the second sign change at $T \simeq 170$ K in Fig. \ref{RhT}(a).
These results show that the details of electron-hole compensation is different between LaSb and LaBi, however, Fig. \ref{RH}(c) suggests that XMR has comparable magnitude in both systems.
Having electron-hole compensation is necessary for large MR as shown in Silver dichalchogenides \cite{xu_large_1997, lee_band-gap_2002} but the detailed degree of compensation does not determine the magnitude of XMR in TSMs. 
Our hypothesis is that the orbital texture on the electron band plays the key role in determining the magnitude of XMR in topological semimetals.
Comparing the mobility of hole pockets from Table \ref{T2} with electron pockets from table \ref{T1} shows that $\mu_h$ is an order of magnitude smaller than $\mu_e$ proving that the quasi-2D electron pockets indeed dominate electrical transport at low temperatures where XMR appears.

\begin{table}
\caption{\label{T2} 
The hole carrier density $n_h$ and mobility $\mu_h$ from the two band fit (Eq. \ref{twob}) are listed for LaSb and LaBi.
}
\begin{ruledtabular}
\begin{tabular}{ccc}
Material & $n_h$ & $\mu_h$ 
\\
         & [cm$^{-3}$] & [cm$^2$V$^{-1}$s$^{-1}$]   
\\       
\colrule 
LaSb & $7.0\times 10^{20}$ & $650$  
\\
LaBi & $2.7\times 10^{20}$ & $780$  

\end{tabular}
\end{ruledtabular}
\end{table}

\section{\label{universal}Triangular Phase Diagram in Topological Semimetals}

\begin{figure}
\includegraphics[width=3.5in]{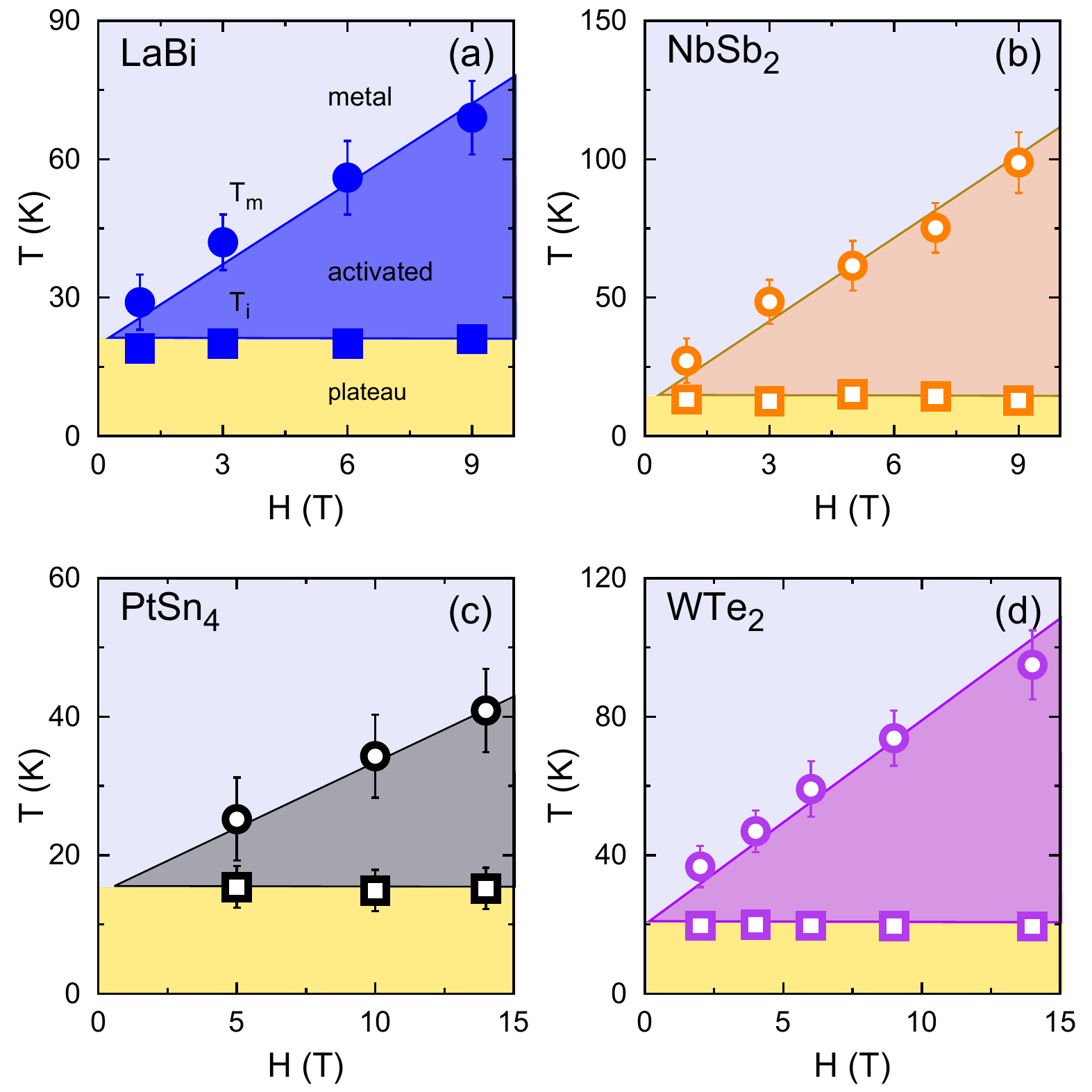}
\caption{\label{UPD} 
(a) Triangular $T$-$H$ phase diagram of LaBi, same as in Fig. \ref{PD}(a), compared to three different topological semimetals in the subsequent panels.
(b) Triangular $T$-$H$ phase diagram in NbSb$_2$ constructed with the data from Ref. \cite{wang_anisotropic_2014}.
\Tm~(circles) and \Ti~(squares) are the temperatures of resistivity minimum and inflection as explained in section \ref{temperature}.
Empty symbols are used for all the adapted data. 
(c) Triangular $T$-$H$ phase diagram in PtSn$_4$ constructed with the data from Ref. \cite{mun_magnetic_2012}.
(d) Triangular $T$-$H$ phase diagram in WTe$_2$ constructed wit the data from Ref. \cite{ali_large_2014}.
Aside from variations in the absolute values of \Tm~and \Ti~which depend on sample purity and XRM magnitude, these different topological semimetals share the same phase diagram. 
}
\end{figure}

Our results in LaSb and LaBi can be summarized as below.\\
(1) $\rho(T)$ at zero field shows a nearly perfect metal with very small $\rho_0$ (Fig. \ref{RT}).\\
(2) $\rho(T)$ in field shows a particular profile with a field-induced activation at \Tm~ and a plateau below \Ti~(Fig. \ref{RT}).\\
(3) The field dependence of \Tm~and \Ti~constructs a triangular phase diagram where \Tm~and \Ti~diverge with increasing field and converge with decreasing field (Fig. \ref{PD}).\\
(4) The field-induced activation of resistivity results in extreme magnetoresistance (XMR) that correlates with RRR (Fig. \ref{RH}).\\
(5) The angle dependence of SdH oscillations shows that quasi-2D Fermi surfaces dominate electrical transport at low temperatures (Fig. \ref{Angle}).\\
(6) From the band structure, these quasi-2D surfaces have a mixed $d$-$p$ orbital texture due to spin-orbit coupling (Fig. \ref{Calc}).
XMR is possibly the consequence of disturbing such orbital texture by a magnetic field.\\
(7) The temperature and the field dependence of Hall effect show multi-band characteristics. 
A better electron-hole compensation is observed in LaBi compared to LaSb (Fig. \ref{RhT}), however, XMR is comparable between the two compounds as shown in Fig. \ref{RH}(c).
These results suggest that XMR in LaSb and LaBi originate from the mixed orbital texture of their quasi-2D Fermi surfaces.
In Appendix \ref{DFTs} we show that a similar orbital texture exists in the Fermiology of other topological semimetals with XMR including NbSb$_2$, PtSn$_4$, and WTe$_2$ (Fig. \ref{Calc_All}).
The $\rho(T)$ and $\rho(H)$ profiles of these seemingly different materials are quite similar with a resistivity minimum at \Tm~and inflection at \Ti.
From the existing transport data in NbSb$_2$ \cite{wang_anisotropic_2014}, PtSn$_4$ \cite{mun_magnetic_2012}, and WTe$_2$ \cite{ali_large_2014}, we extract \Tm~and \Ti, construct their $T$-$H$ phase diagrams, and compare to the triangular phase diagram of LaBi in Fig. \ref{UPD}. 
Remarkably, the same triangular phase diagram is observed despite the chemical and structural differences of these materials.
Figs. \ref{UPD} and \ref{Calc_All} provide compelling evidence to identify orbital mixing on quasi-2D Fermi surfaces as the origin of XMR in TSMs.
%
%


\section*{ACKNOWLEDGMENTS}

We thank 
S.~R.~Julian,
and 
L. Muechler
for helpful discussions.
This research was supported by the Gordon and Betty Moore Foundation under the EPiQS program, grant GBMF 4412. 
S.K.K is supported by the ARO MURI on topological insulators, grant W911NF-12-1-0461.  


\appendix

\section{\label{refinement}Characterization of LaSb and LaBi samples with XRD}

\begin{figure}
\includegraphics[width=3.5in]{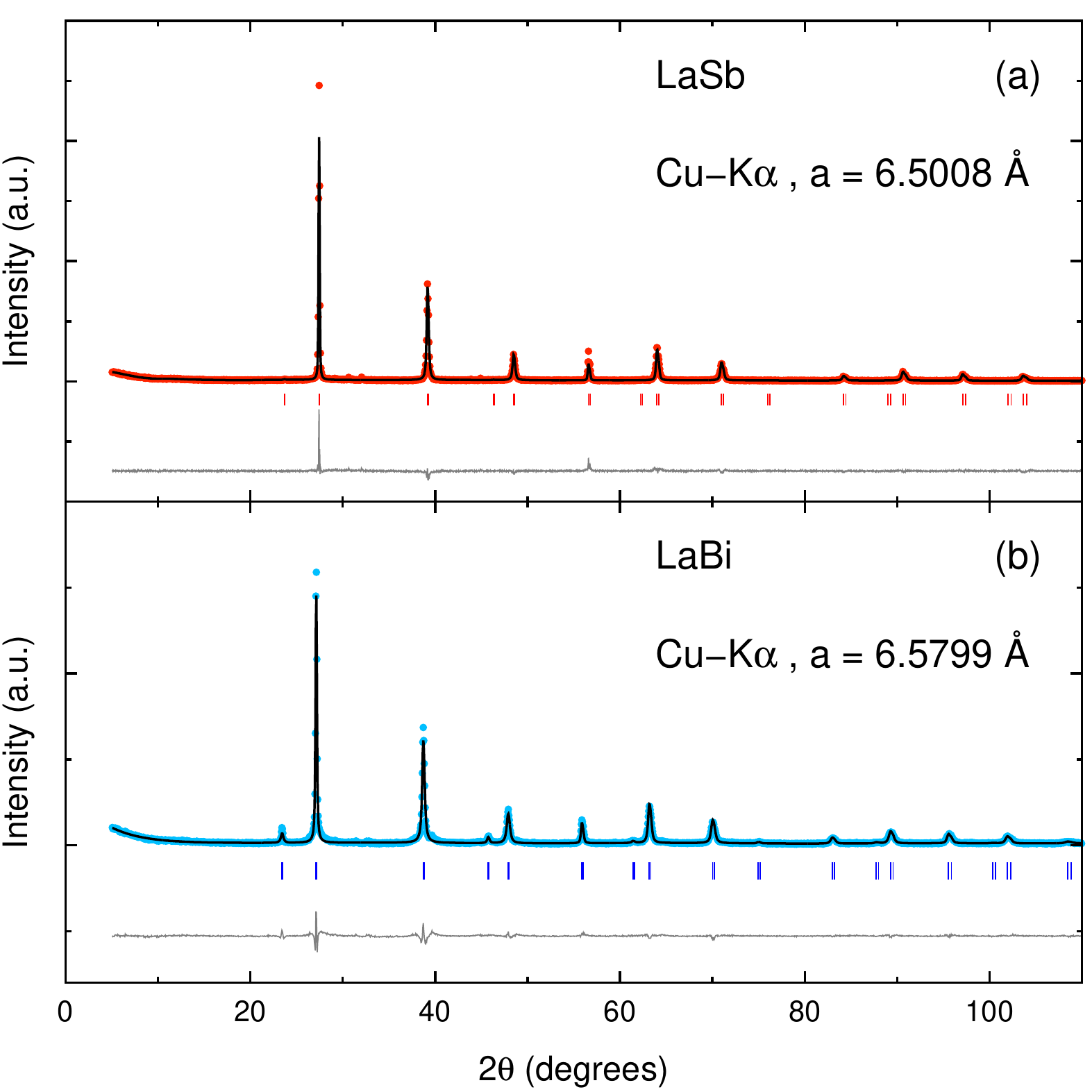}
\caption{\label{XRD} 
(a) Rietveld refinement (solid black line) on powder XRD pattern (red circles) for LaSb.
Lattice parameter from the refinement is quoted on the figure. 
(b) Rietveld refinement (solid black line) on powder XRD pattern (blue circles) for LaBi. 
}
\end{figure}

Fig \ref{XRD}(a) and \ref{XRD}(b) show powder x-ray diffraction patterns of our LaSb and LaBi samples. 
XRD date were acquired in a Bruker D8 ADVANCE ECO system with LYNXEYE XE high resolution energy-dispersive 1D detector. 
Rietveld refinement of these patterns are shown as solid black lines in Fig. \ref{XRD}.
Refinements were done through the Fullprof software \cite{rodriguez-carvajal_recent_1993} using Thompson-Cox-Hastings pseudo Voight profile convoluted with axial divergence asymmetry. 

\section{\label{DFTs}DFT calculations on NbSb$_2$, PtSn$_4$, and WTe$_2$}

\begin{figure}
\includegraphics[width=3.5in]{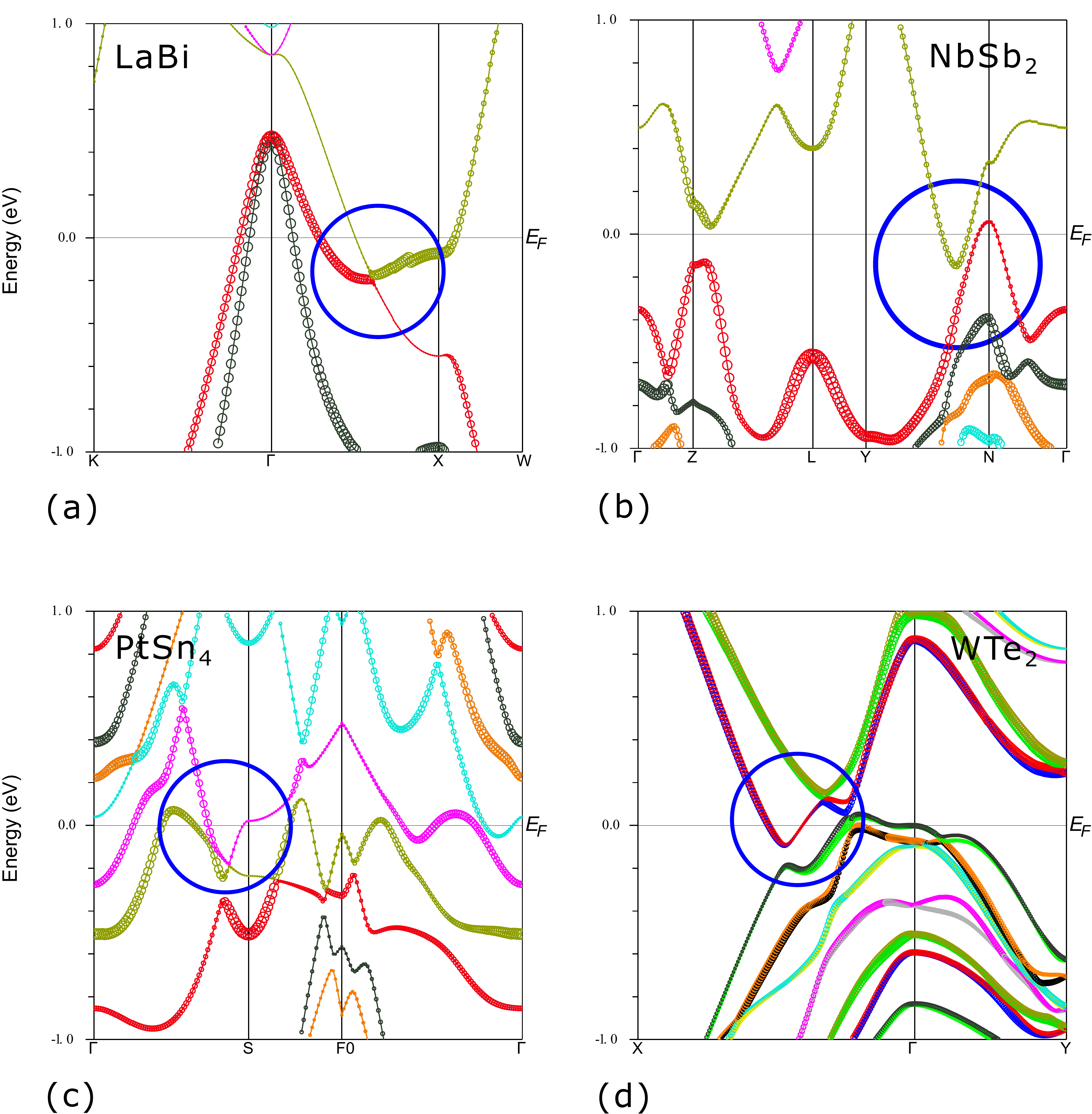}
\caption{\label{Calc_All} 
(a) Band structure of LaBi, same as in Fig. \ref{Calc}(b). 
The blue circle marks the region of $d$-$p$ orbital mixing.
(b) Band structure of NbSb$_2$ with a similar $d$-$p$ mixing inside the blue circle.
(c) Band structure of PtSn$_4$ with similar $d$-$p$ mixing.
(d) Band structure of WTe$_2$ with similar $d$-$p$ mixing.
}
\end{figure}

Fig. \ref{Calc_All} shows the results of our density functional theory (DFT) calculations using WIEN2k code \cite{blaha_wien2k_2001} on LaBi, NbSb$_2$, PtSn$_4$, and WTe$_2$.
Regions of mixed $d$-$p$ orbital texture due to spin-orbit coupling are marked with blue circles. 
All these semimetals have quasi-2D Fermi surfaces similar to Fig. \ref{Calc}(c) and \ref{Calc}(d) for LaSb and LaBi \cite{wang_anisotropic_2014, mun_magnetic_2012, ali_large_2014}.
Extreme magnetoresistance in these materials has the same triangular phase diagram as in lanthanum monopnictides (Fig. \ref{UPD}).
Similar phase diagram (Fig. \ref{UPD}) and similar band structure (Fig. \ref{Calc_All}) in these materials point towards a common origin for XMR in TSMs as discussed in section \ref{universal}.



\bibliography{LaBi_22dec2015}

\end{document}